\documentclass[12pt]{article}
\usepackage{amsmath,amsfonts,amssymb}
\usepackage{graphicx}

      \usepackage{epstopdf}

\makeatletter
\def\numberbysection{\@addtoreset{equation}{section}
    \def\theequation{\thesection.\arabic{equation}}}
\makeatother

\numberbysection

\newcommand{\be}{\begin{eqnarray}}
\newcommand{\ee}{\end{eqnarray}}

\newcommand{\ads}{{\rm AdS}_3/{\rm CFT}_2}

\begin{document}

\begin{titlepage}
\vspace{.5in}
\begin{center}

\LARGE Exact $S$-matrices for $\ads$\\
\vspace{1in}
\large Changrim Ahn \footnote{
       Department of Physics, Ewha Womans University,
       Seoul 120-750, South Korea; ahn@ewha.ac.kr} 
and Diego Bombardelli \footnote{Centro de F\'isica do Porto, Departamento de F\'isica e Astronomia, Faculdade de Ci\^encias da Universidade do Porto, Rua do Campo Alegre 687, 4169-007 Porto, Portugal; diego.bombardelli@fc.up.pt}\\ 

\end{center}

\vspace{.5in}

\begin{abstract}
We propose exact $S$-matrices for the $\ads$ duality 
between Type IIB strings on ${\rm AdS}_3\times S^3\times M_4$
with $M_4=S^3\times S^1$ or $T^4$ and the corresponding two-dimensional
conformal field theories.
We fix the complete
two-particle $S$-matrices for both those cases of $\ads$, on the basis of the symmetries $su(1|1)$ and $su(1|1)\times su(1|1)$, respectively preserved by their vacua.
A crucial justification comes from the derivation of the all-loop
Bethe ansatz matching exactly the recent conjecture proposed by
\cite{Babichenko:2009dk} and \cite{OhlssonSax:2011ms}.
\end{abstract}

\end{titlepage}

\setcounter{footnote}{0}

\section{Introduction}\label{sec:intro}

The discovery of integrable structures on both sides of the ${\rm AdS}_5/{\rm CFT}_4$ correspondence \cite{Minahan:2002ve, Bena:2003wd}, was crucial in understanding and determine exactly some important physical quantities (see \cite{Beisert:2010jr} and references therein) in ${\cal N}=4$ Super-Yang-Mills and the IIB superstring theory on ${\rm AdS}_5\times S^5$ in the planar limit. 

From the integrability point of view, one of the most recently investigated example of such gauge/string duality is the ${\rm AdS}_3/{\rm CFT}_2$ correspondence between IIB superstring theory on ${\rm AdS}_3 \times S^3 \times S^3 \times S^1$ or ${\rm AdS}_3 \times S^3 \times T^4$ backgrounds with RR fluxes and yet quite unknown two-dimensional CFTs \footnote{In the case of ${\rm AdS}_3 \times S^3 \times T^4$, the CFT dual is a ${\cal N}=(4,4)$ theory on a symmetric product of $T^4$.}. Indeed, while the NS ${\rm AdS}_3/{\rm CFT}_2$ was solved completely by implementing techniques typical of two-dimensional CFTs \cite{Giveon:1998ns}, the RR counterpart remains quite obscure: there the usual 2D CFT methods fail, then one of the most promising way to attack this problem is given by integrability techniques. 

This investigation started in \cite{Babichenko:2009dk, OhlssonSax:2011ms}, where a set of all-loop Bethe equations, describing in principle at any coupling the asymptotic spectrum of the string energies and the dimensions of the yet unknown gauge operators, were proposed on the basis of classical integrability of the corresponding supercoset sigma models \footnote{Some first tests of these Bethe equations against string energy calculations have been performed in \cite{Rughoonauth:2012qd}.}. Unfortunately this approach cannot take into account the contribution of some (massless) modes of the full string theory. Some progress in the direction of incorporating them has been done very recently in \cite{Sax:2012jv}, where a set of Bethe equations, completely decoupled from the others, has been proposed in order to describe the massless modes. Now, this would mean that the $S$-matrix between the massless and massive modes become trivial and lead to independent set of Bethe equations. 

The aim of this Letter is to propose an $S$-matrix for the massive modes, in order to derive, on a firmer ground, the Bethe equations proposed in \cite{Babichenko:2009dk, OhlssonSax:2011ms}. We shall do this by using an analytic Bethe ansatz involving transfer matrix eigenvalues derived from the diagonalization of $su(1|1)$ and $su(1|1)\times su(1|1)$ invariant $S$-matrices, respectively for the ${\rm AdS}_3 \times S^3 \times S^3 \times S^1$ and ${\rm AdS}_3 \times S^3 \times T^4$ cases.
Basically, this means that the $S^3 \times S^1$ ($T^4$) case has total symmetry $d(2,1;\alpha)\times d(2,1;\alpha)$ ($psu(1,1|2) \times psu(1,1|2)$), but  the on-shell particle symmetry which preserves the vacuum is $su(1|1)$ ($su(1|1)\times su(1|1)$).

In the different context of open ${\rm AdS}_5/{\rm CFT}_4$ spin chains, the analytic Bethe ansatz built on an $su(1|1)$-invariant $S$-matrix, previously found in \cite{Beisert:2005fw, Beisert:2005wm}, was already performed, without considering possible scalar factors, by \cite{RafEri} in order to determine the corresponding transfer matrix eigenvalues and Bethe equations.
On the other hand, an $su(1|1)\times su(1|1)$-invariant $S$-matrix was proposed in \cite{David:2010yg} to describe the scattering of magnons in ${\rm AdS}_3 \times S^3 \times T^4$; however, only magnons \footnote{Which were previously argued in \cite{David:2008yk} being BPS states in a centrally extended $\mathfrak{su}(1|1)\times \mathfrak{su}(1|1)$ algebra.} in the $su(2)$ sector were analyzed there, in order to derive the dressing phase up to one-loop, and Bethe equations were not derived.

\section{Spectrum and $S$-matrix}

\subsection{${\rm AdS}_3\times S^3\times T^4$}
For the case of ${\rm AdS}_3\times S^3\times T^4$, the spectrum consists of
eight massive modes whose energy-momentum dispersion relation 
is given by
\be
E = \sqrt{\frac{1}{4}+4 h^{2}(\lambda) \sin^{2}\frac{p}{2}} \,,
\ee
where $h$ is an almost unknown function of the 't Hooft coupling $\lambda$: its strong coupling behavior has been predicted to be $h(\lambda)\simeq\sqrt{\lambda}/2\pi$ by \cite{Babichenko:2009dk}, while the one-loop correction has been calculated recently by \cite{Sundin:2012gc}.
These are grouped into bifundamentals of
$su(1|1)\times su(1|1)$, which we refer to ``A'' and ``B''. 
The $S$-matrices among these bifundamentals are given by tensor
products of two $su(1|1)$-invariant $S$-matrices as follows: 
\be
S^{(AA)}(p_1,p_2)&=& S^{(BB)}(p_1,p_2) 
=S_0(p_1,p_2) \left[{\hat S}(p_1,p_2)\otimes 
{\hat S}(p_1,p_2)\right],\\
S^{(AB)}(p_1,p_2)&=& S^{(BA)}(p_1,p_2)
={\tilde S}_0(p_1,p_2) \left[{\hat S}(p_1,p_2)\otimes
{\hat S}(p_1,p_2)\right],
\ee
where \cite{Beisert:2005wm, RafEri, David:2010yg}
\be
{\hat S}(p_1,p_2)=
\left( \begin{array}{cccc}
1 & 0 & 0 & 0 \\
0 & \frac{x^-_1-x^-_2}{x^+_1-x^-_2} & \frac{x^+_1-x^-_1}{x^+_1-x^-_2}\frac{\omega_2}{\omega_1}  & 0 \\
0 & \frac{x^+_2-x^-_2}{x^+_1-x^-_2}\frac{\omega_1}{\omega_2}   & \frac{x^+_1-x^+_2}{x^+_1-x^-_2}  & 0 \\
0 & 0 & 0 & \frac{x^-_1-x^+_2}{x^+_1-x^-_2}  
\end{array} \right) 
\label{su11Smatrix}
\ee
and we set $\omega_{1,2}=\omega(p_{1,2})=1$. The $x^{\pm}$ variables are the usual Zhukowski variables defined by 
\be
\frac{x^+}{x^-}=e^{ip}, x^{+}+\frac{1}{x^{+}}-x^{-}-\frac{1}{x^{-}}=\frac{i}{h(\lambda)}.
\ee
 
The $S$-matrix (\ref{su11Smatrix}) satisfies the unitarity condition, but it does not have crossing symmetry \cite{RafEri}. 
An attempt to derive the crossing symmetry relations for the $\mathfrak{su}(1|1)$ algebra has been put forward in \cite{David:2010yg} by using the antipode operation, but this implies, in this case, a transformation on the kinematic variables ($x^{\pm}\rightarrow x^{\mp}$) that does not correspond to the particle-antiparticle transformation ($x^{\pm}\rightarrow 1/x^{\pm}$).
Then we guess a possible expression for the scalar factors on the basis of the unitarity and the final matching with the Bethe equations proposed by \cite{Babichenko:2009dk, OhlssonSax:2011ms}:
\be
S_0(p_1,p_2)=\frac{x^+_1-x^-_2}{x^-_1-x^+_2}\frac{1-\frac{1}{x^+_1 x^-_2}}{1-\frac{1}{x^-_1 x^+_2}}
\sigma^2(p_1,p_2)\frac{x_1^-}{x_1^+}\frac{x_2^+}{x_2^-},\quad
{\tilde S}_0(p_1,p_2)=\sigma^{-2}(p_1,\bar p_2)\frac{x_1^+}{x_1^-}\frac{x_2^+}{x_2^-},
\label{S0}
\ee
where $\sigma(p_1,p_2)$ is the BES dressing phase \cite{Beisert:2006ez} and $\bar p$ denotes the momentum of an antiparticle, such that $x^{\pm}(\bar p)=1/x^{\pm}(p)$.
The scalar factors (\ref{S0}) satisfy the relation $S_0(p_1,p_2)=S_0(\bar p_1,\bar p_2), \tilde S_0(p_1,p_2)=\tilde S_0(\bar p_1,\bar p_2)$, that will be important later for the construction of the Bethe equations, and unitarity:
\be
&&\hspace{-0.8cm}S_0(p_1,p_2)S_0(p_2,p_1)=S_{su(2)}(p_1,p_2)\sigma^2(p_1,p_2)\frac{x_1^-}{x_1^+}\frac{x_2^+}{x_2^-}S_{su(2)}^{-1}\sigma^{-2}(p_1,p_2)\frac{x_2^-}{x_2^+}\frac{x_1^+}{x_1^-}=1,
\label{unit1}\\
&&\hspace{-0.8cm}\tilde S_0(p_1,p_2)\tilde S_0(p_2,p_1)=\sigma^{-2}(p_1,\bar p_2)\frac{x_1^+}{x_1^-}\frac{x_2^+}{x_2^-}\sigma^{-2}(\bar p_2,p_1)\frac{x_1^-}{x_1^+}\frac{x_2^-}{x_2^+}
=1\,,
\label{unit2}
\ee
where $S_{su(2)}(p_1,p_2)=\frac{x^+_1-x^-_2}{x^-_1-x^+_2}\frac{1-\frac{1}{x^+_1 x^-_2}}{1-\frac{1}{x^-_1 x^+_2}}$. 

The Bethe-Yang equations are derived from a periodic boundary condition (PBC).
On a circle with circumference $L$, we put $N_A$ number of ``A'' particles
with momenta $\{p^A_1,p^A_2,\ldots,p^A_{N_A}\}$ and
$N_B$ number of ``B'' particles
with momenta $\{p^B_1,p^B_2,\ldots,p^B_{N_B}\}$.
Now we choose an ``A'' particle with a momentum $p^A_j$ and move it around
the circle by scattering with all the other particles and similarly for a 
``B'' particle with a momentum $p^B_j$.
Since this virtual process does not change any configuration, we arrive at
PBC conditions
\be
e^{ip^A_jL}&=&\prod_{k=1,\neq j}^{N_A}S_0(p^A_j,p^A_k)
\prod_{k=1}^{N_B}{\tilde S}_0(p^A_j,p^B_k)
\left[{\widehat T}_{su(1|1)}(p^A_j|\{p^A_l,p^B_l\})
\otimes {\widehat T}_{su(1|1)}(p^A_j|\{p^A_l,p^B_l\})\right],
\label{T4PBCi}\nonumber\\
e^{ip^B_jL}&=&\prod_{k=1,\neq j}^{N_B}S_0(p^B_j,p^B_k)
\prod_{k=1}^{N_A}{\tilde S}_0(p^B_j,p^A_k)
\left[{\widehat T}_{su(1|1)}(p^B_j|\{p^A_l,p^B_l\})
\otimes {\widehat T}_{su(1|1)}(p^B_j|\{p^A_l,p^B_l\})\right],\nonumber
\label{T4PBCii}
\ee
where ${\widehat T}_{su(1|1)}$ is a transfer matrix made of the 
$su(1|1)$-invariant $S$-matrix,
\be
{\widehat T}_{su(1|1)}(p|\{p^A_l\},\{p^B_l\})=
{\rm str}_{a}\left[{\hat S}_{aA_1}(p,p^A_1) \cdots
{\hat S}_{aA_{N_A}}(p,p^A_{N_A})
{\hat S}_{aB_1}(p,p^B_1) \cdots
{\hat S}_{aB_{N_B}}(p,p^B_{N_B})\right],
\label{transfer}
\ee
and $a$, $A_i$ and $B_i$ stand for a two-dimensional vector space which
the $S$-matrices act on.

\subsection{${\rm AdS}_3\times S^3\times S^3\times S^1$}
The spectrum of ${\rm AdS}_3\times S^3\times S^3\times S^1$ is a bit more complicated.
Denoting $l, R_1, R_2$ the radii of ${\rm AdS}_3$ and the two $S^3$'s respectively, one has the following  relation
\be
\frac{1}{R_1^2}+\frac{1}{R_2^2}=\frac{1}{l^2}.
\ee
By defining $\alpha=l^2/R_1^2$, one can find two massive multiplets,
each of which consists of two bosons and two fermions, with two 
different masses:
\be
E_l = \sqrt{m_l^2+4 h^{2}(\lambda) \sin^{2}\frac{p}{2}} \,,
\quad l=1,3\,,
\ee
where
\be
m_1=\alpha,\quad m_3=1-\alpha\,.
\ee

We propose that the four particles with mass $m_1$ are grouped into two fundamentals of $su(1|1)$,
which we refer to ``$1$'' and ``${\bar 1}$''; and similarly the other four particles with mass $m_3$ 
into two additional fundamentals of $su(1|1)$, which we refer to ``$3$'' and ``${\bar 3}$''.
In this case the Zhukowsky variables are defined as \cite{OhlssonSax:2011ms}:
\be
x^{+}_{1,\bar1}+\frac{1}{x^{+}_{1,\bar1}}-x^{-}_{1,\bar1}-\frac{1}{x^{-}_{1,\bar1}}=\frac{i\alpha}{h(\lambda)}\,;\quad x^{+}_{3,\bar3}+\frac{1}{x^{+}_{3,\bar3}}-x^{-}_{3,\bar3}-\frac{1}{x^{-}_{3,\bar3}}=\frac{i(1-\alpha)}{h(\lambda)}.
\ee
The $S$-matrices among these four doublets are given by single $su(1|1)$-invariant $S$-matrices 
as follows:
\be
S^{(11)}(p_1,p_2)&=& S^{(33)}(p_1,p_2)=S^{({\bar 1}{\bar 1})}(p_1,p_2)=S^{({\bar 3}{\bar 3})}(p_1,p_2)
=S_0(p_1,p_2) {\hat S}(p_1,p_2)\\
S^{(1{\bar 1})}(p_1,p_2)&=& S^{(3{\bar 3})}(p_1,p_2)=
S^{({\bar 1}1)}(p_1,p_2)=S^{({\bar 3}3)}(p_1,p_2)
={\tilde S}_0(p_1,p_2) {\hat S}(p_1,p_2)\\
S^{(13)}(p_1,p_2)&=& S^{(31)}(p_1,p_2)=S^{({\bar 1}{\bar 3})}(p_1,p_2)=S^{({\bar 3}{\bar 1})}(p_1,p_2)
={\hat S}(p_1,p_2)\\
S^{(1{\bar 3})}(p_1,p_2)&=& S^{(3{\bar 1})}(p_1,p_2)=
S^{({\bar 3}1)}(p_1,p_2)= S^{({\bar 1}3)}(p_1,p_2)
={\hat S}(p_1,p_2),
\ee
where ${\hat S}(p_1,p_2)$ is given in Eq.(\ref{su11Smatrix}) and the scalar factors $S_0$ and ${\tilde S}_0$ are defined in Eq.(\ref{S0}).

The Bethe-Yang equations can be written in a similar way as before.
On a circle with circumference $L$, we put $N_1$ number of ``$1$'' particles
with momenta $\{p^{1}_1,p^1_2,\ldots,p^1_{N_1}\}$,
$N_{\bar 1}$ number of ``${\bar 1}$'' particles
with momenta $\{p^{\bar 1}_1,p^{\bar 1}_2,\ldots,p^{\bar 1}_{N_{\bar 1}}\}$,
$N_3$ number of ``$3$'' particles
with momenta $\{p^{3}_1,p^3_2,\ldots,p^3_{N_3}\}$, and
$N_{\bar 3}$ number of ``${\bar 3}$'' particles
with momenta $\{p^{\bar 3}_1,p^{\bar 3}_2,\ldots,p^{\bar 3}_{N_{\bar 3}}\}$.
From these configuration, the PBC equations become
\be
e^{ip^{1}_jL}&=&\prod_{k=1,\neq j}^{N_1}S_0(p^{1}_j,p^{1}_k)
\prod_{k=1}^{N_{\bar 1}}{\tilde S}_0(p^{1}_j,p^{\bar 1}_k)
\cdot
{\widehat T}_{su(1|1)}(p^1_j|\{p^1_l,p^{\bar 1}_l,p^3_l,p^{\bar 3}_l\})
\label{S3PBCi}\\
e^{ip^{\bar 1}_jL}&=&\prod_{k=1,\neq j}^{N_{\bar 1}}S_0(p^{\bar 1}_j,p^{\bar 1}_k)
\prod_{k=1}^{N_{1}}{\tilde S}_0(p^{\bar 1}_j,p^{1}_k)
\cdot
{\widehat T}_{su(1|1)}(p^{\bar 1}_j|\{p^1_l,p^{\bar 1}_l,p^3_l,p^{\bar 3}_l\})
\label{S3PBCii}\\
e^{ip^{3}_jL}&=&\prod_{k=1,\neq j}^{N_1}S_0(p^{3}_j,p^{3}_k)
\prod_{k=1}^{N_{\bar 3}}{\tilde S}_0(p^{3}_j,p^{\bar 3}_k)
\cdot
{\widehat T}_{su(1|1)}(p^3_j|\{p^1_l,p^{\bar 1}_l,p^3_l,p^{\bar 3}_l\})
\label{S3PBCiii}\\
e^{ip^{\bar 3}_jL}&=&\prod_{k=1,\neq j}^{N_{\bar 3}}S_0(p^{\bar 3}_j,p^{\bar 3}_k)
\prod_{k=1}^{N_{3}}{\tilde S}_0(p^{\bar 3}_j,p^{3}_k)
\cdot
{\widehat T}_{su(1|1)}(p^{\bar 3}_j|\{p^1_l,p^{\bar 1}_l,p^3_l,p^{\bar 3}_l\}),
\label{S3PBCiv}
\ee
where ${\widehat T}_{su(1|1)}$ is given in Eq.(\ref{transfer}).

\section{Derivation of asymptotic Bethe ansatz equations}

\subsection{Diagonalization of the transfer matrix}

The $su(1|1)$ transfer matrix has been diagonalized by the analytic Bethe ansatz method in \cite{RafEri}.
The eigenvalues can be expressed by
\be
\Lambda(p|\{p_\ell\},\{\lambda_j\})&=&\Lambda_0(p|\{p_\ell\})A(p|\{\lambda_j\}),\\
\Lambda_0(p|\{p_\ell\})&=&
1-\prod_{\ell=1}^{N}\left(\frac{x^+(p)-x^+(p_\ell)}{x^+(p)-x^-(p_\ell)}\right),\\
A(p|\{\lambda_j\})&=&\prod_{j=1}^{M}\left(\frac{x^-(p)-x^+(\lambda_j)}{x^+(p)-x^+(\lambda_j)}\right),
\ee
and the magnonic variables $\lambda_j$ satisfy
\be
1=\prod_{\ell=1}^{N}\left(\frac{x^+(\lambda_j)-x^-(p_\ell)}{x^+(\lambda_j)-x^+(p_\ell)}\right).
\ee
Here, we have used a short notation that $N=N_A+N_B$ and $\{p_\ell\}=\{p^A_l,p^B_l\}$ for ${\rm AdS}_3\times S^3\times T^4$; $N=N_1+N_{\bar 1}+N_3+N_{\bar 3}$, $\{p_\ell\}=\{p^1_l,p^{\bar 1}_l,p^3_l,p^{\bar 3}_l\}$ for ${\rm AdS}_3\times S^3\times S^3\times S^1$, respectively.

Inserting these into Eqs.(\ref{T4PBCi}-\ref{T4PBCii}), we get the asymptotic Bethe ansatz equations for 
${\rm AdS}_3\times S^3\times T^4$:
\be
e^{ip^A_jL}&=&\prod_{k=1,\neq j}^{N_A}S_0(p^A_j,p^A_k)
\prod_{k=1}^{N_B}{\tilde S}_0(p^A_j,p^B_k)\nonumber\\
&\times&\prod_{j=1}^{M}\left(\frac{x^-(p^A_j)-x^+(\lambda_j)}{x^+(p^A_j)-x^+(\lambda_j)}\right)
\prod_{j=1}^{\overline M}\left(\frac{x^-(p^A_j)-x^+({\overline\lambda}_j)}{x^+(p^A_j)-x^+({\overline \lambda}_j)}\right),
\label{T4BAEi}\\
e^{ip^B_jL}&=&\prod_{k=1,\neq j}^{N_B}S_0(p^B_j,p^B_k)
\prod_{k=1}^{N_A}{\tilde S}_0(p^B_j,p^A_k)\nonumber\\
&\times&
\prod_{j=1}^{M}\left(\frac{x^-(p^B_j)-x^+(\lambda_j)}{x^+(p^B_j)-x^+(\lambda_j)}\right)
\prod_{j=1}^{\overline M}\left(\frac{x^-(p^B_j)-x^+({\overline\lambda}_j)}{x^+(p^B_j)-x^+({\overline \lambda}_j)}\right),
\label{T4BAEii}\\
1&=&\prod_{l=1}^{N_A}\left(\frac{x^+(\lambda_j)-x^-(p^A_l)}{x^+(\lambda_j)-x^+(p^A_l)}\right)
\prod_{l=1}^{N_B}\left(\frac{x^+(\lambda_j)-x^-(p^B_l)}{x^+(\lambda_j)-x^+(p^B_l)}\right),
\label{T4BAEiii}\\
1&=&\prod_{l=1}^{N_A}\left(\frac{x^+({\overline\lambda}_j)-x^-(p^A_l)}{x^+({\overline\lambda}_j)-x^+(p^A_l)}\right)
\prod_{l=1}^{N_B}\left(\frac{x^+({\overline\lambda}_j)-x^-(p^B_l)}{x^+({\overline\lambda}_j)-x^+(p^B_l)}\right).
\label{T4BAEiv}
\ee
\begin{figure}
\begin{centering}
\includegraphics[height=4cm]{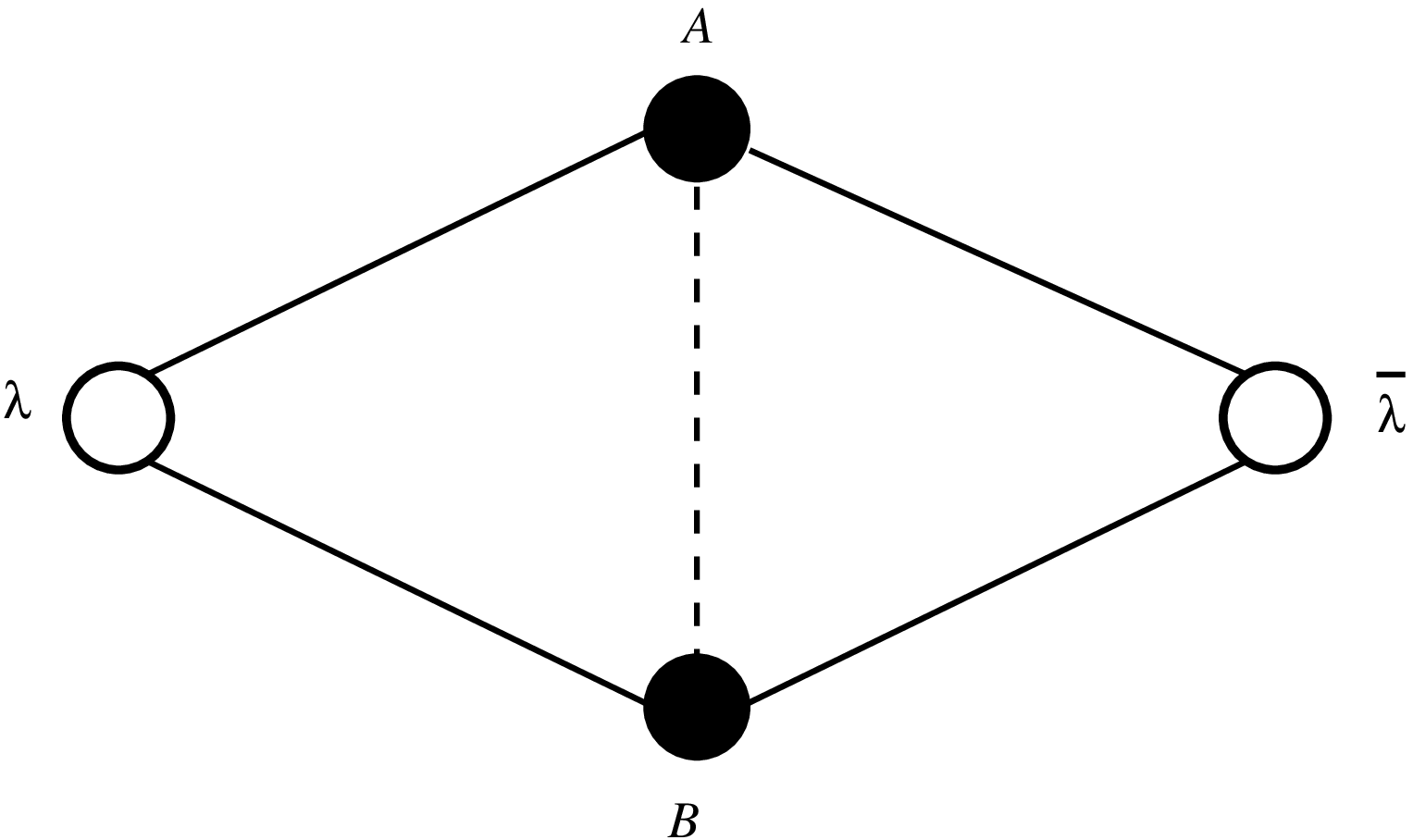}
\par\end{centering}
\caption{${\rm AdS}_3\times S^3\times T^4$: two momentum-carrying nodes (black dots) are connected to
two magnonic nodes (circle).}
\label{fig1}
\end{figure}
This can be represented pictorially by Fig.1.

Similarly, from Eqs.(\ref{S3PBCi}-\ref{S3PBCiv}), we obtain the asymptotic Bethe ansatz equations for 
${\rm AdS}_3\times S^3\times S^3\times S^1$:
\be
e^{ip^{1}_jL}&=&\prod_{k=1,\neq j}^{N_1}S_0(p^{1}_j,p^{1}_k)
\prod_{k=1}^{N_{\bar 1}}{\tilde S}_0(p^{1}_j,p^{\bar 1}_k)
\prod_{j=1}^{M}\left(\frac{x^-(p^1_j)-x^+(\lambda_j)}{x^+(p^1_j)-x^+(\lambda_j)}\right),
\label{S3BAEi}\\
e^{ip^{\bar 1}_jL}&=&\prod_{k=1,\neq j}^{N_{\bar 1}}S_0(p^{\bar 1}_j,p^{\bar 1}_k)
\prod_{k=1}^{N_{1}}{\tilde S}_0(p^{\bar 1}_j,p^{1}_k)
\prod_{j=1}^{M}\left(\frac{x^-(p^1_j)-x^+(\lambda_j)}{x^+(p^1_j)-x^+(\lambda_j)}\right),
\label{S3BAEii}\\
e^{ip^{3}_jL}&=&\prod_{k=1,\neq j}^{N_1}S_0(p^{3}_j,p^{3}_k)
\prod_{k=1}^{N_{\bar 3}}{\tilde S}_0(p^{3}_j,p^{\bar 3}_k)
\prod_{j=1}^{M}\left(\frac{x^-(p^1_j)-x^+(\lambda_j)}{x^+(p^1_j)-x^+(\lambda_j)}\right),
\label{S3BAEiii}\\
e^{ip^{\bar 3}_jL}&=&\prod_{k=1,\neq j}^{N_{\bar 3}}S_0(p^{\bar 3}_j,p^{\bar 3}_k)
\prod_{k=1}^{N_{3}}{\tilde S}_0(p^{\bar 3}_j,p^{3}_k)
\prod_{j=1}^{M}\left(\frac{x^-(p^1_j)-x^+(\lambda_j)}{x^+(p^1_j)-x^+(\lambda_j)}\right),
\label{S3BAEiv}\\
1&=&\prod_{\ell=1}^{N}\left(\frac{x^+(\lambda_j)-x^-(p_\ell)}{x^+(\lambda_j)-x^+(p_\ell)}\right).
\label{S3BAEv}
\ee
\begin{figure}
\begin{centering}
\includegraphics[height=4cm]{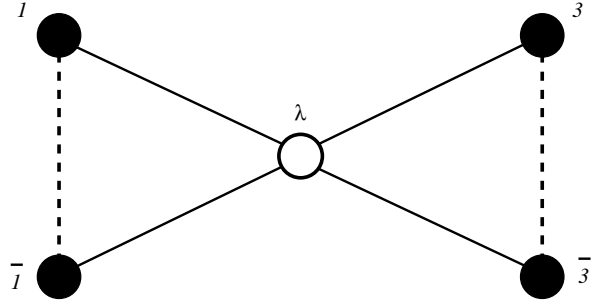}
\par\end{centering}
\caption{${\rm AdS}_3\times S^3\times S^3\times S^1$: four momentum-carrying nodes (black dots) are connected to a single magnonic node (circle).}
\label{fig2}
\end{figure}
These sets of Bethe ansatz equations can be represented pictorially by Fig.2.

\subsection{Comparison to the Bethe equations of \cite{Babichenko:2009dk, OhlssonSax:2011ms}}

In order to translate Eqs.(\ref{T4BAEi}-\ref{T4BAEiv}) into the notation of \cite{Babichenko:2009dk, OhlssonSax:2011ms}, we have to replace $p_B$ by $\bar p_B$ and disentangle the two 
``magnonic'' variables into four.
In Eq.(\ref{T4BAEi}), for instance, the first step involves the second factor:
\be
\prod_{k=1}^{N_B}{\tilde S}_0(p^A_j,\bar p^B_k)=
\prod_{k=1}^{N_B}\sigma^{-2}(p_j^A,p_k^B)\frac{x_j^+}{x_j^-}\frac{x_k^-}{x_k^+}\,.
\ee 
Now, since from the momentum constraint we have that $\prod_{k=1}^{N_A}\frac{x_k^+}{x_k^-}\prod_{k=1}^{N_B}\frac{x_k^-}{x_k^+}=1$, and $\frac{x_j^+}{x_j^-}=e^{ip_j}$, finally we get, ignoring for the moment the magnonic part (setting to zero both $M$ and $\overline M$):
\be
e^{ip_j^A(L+N_A-N_B)}=\prod_{k=1,\neq j}^{N_A}\frac{x^+_j-x^-_k}{x^-_j-x^+_k}\frac{1-\frac{1}{x^+_j x^-_k}}{1-\frac{1}{x^-_j x^+_k}}
\sigma^2(p_j^A,p_k^A)\prod_{k=1}^{N_B}\sigma^{-2}(p_j^A,p_k^B)\,.
\ee
In the case of Eq.(\ref{T4BAEii}), we get:
\be
e^{-ip_j^B(L+N_B-N_A)}=\prod_{k=1,\neq j}^{N_B}\frac{x^+_j-x^-_k}{x^-_j-x^+_k}\frac{1-\frac{1}{x^+_j x^-_k}}{1-\frac{1}{x^-_j x^+_k}}
\sigma^2(p_j^B,p_k^B)\prod_{k=1}^{N_A}\sigma^{-2}(p_j^B,p_k^A)\,.
\ee

Now, in order to complete the comparison, we need also to redefine the 
magnonic variables (After this, Fig.1  changes to  Fig.3.):
\be
x^+(\lambda_j)=x_{1,j};\quad j=1,\dots K_1;\quad x^+(\lambda_{K_1+j})=1/x_{\bar1,j};\quad j=1,\dots K_{\bar1};\quad M=K_1+K_{\bar1}\\
x^+(\overline\lambda_j)=x_{3,j};\quad j=1,\dots K_3;\quad x^+(\overline\lambda_{K_3+j})=1/x_{\bar3,j};\quad j=1,\dots K_3;\quad M=K_3+K_{\bar3}
\ee
Then the Eqs.(\ref{T4BAEi}-\ref{T4BAEiv}) become:
\be
&&e^{ip^A_j(L+K_A-K_B+K_{\bar1}+K_{\bar3})}=\prod_{k=1,\neq j}^{K_A}\frac{x^+_j-x^-_k}{x^-_j-x^+_k}\frac{1-\frac{1}{x^+_j x^-_k}}{1-\frac{1}{x^-_j x^+_k}}
\sigma^2(p_j^A,p_k^A)\prod_{k=1}^{N_B}\sigma^{-2}(p_j^A,p_k^B)\nonumber\\
&&\times\prod_{j=1}^{K_1}\frac{x^-(p^A_j)-x_{1,j}}{x^+(p^A_j)-x_{1,j}}\prod_{j=1}^{K_{\bar1}}\frac{1-\frac{1}{x^-(p^A_j)x_{\bar1,j}}}{1-\frac{1}{x^+(p^A_j)x_{\bar1,j}}}
\prod_{j=1}^{K_3}\frac{x^-(p^A_j)-x_{3,j}}{x^+(p^A_j)-x_{3,j}}\prod_{j=1}^{K_{\bar3}}\frac{1-\frac{1}{x^-(p^A_j)x_{\bar3,j}}}{1-\frac{1}{x^+(p^A_j)x_{\bar3,j}}}\,,
\\
&&e^{-ip^B_j(L+K_A-K_B+K_{\bar1}+K_{\bar3})}=\prod_{k=1,\neq j}^{K_B}\frac{x^+_j-x^-_k}{x^-_j-x^+_k}\frac{1-\frac{1}{x^+_j x^-_k}}{1-\frac{1}{x^-_j x^+_k}}
\sigma^2(p_j^B,p_k^B)\prod_{k=1}^{N_A}\sigma^{-2}(p_j^B,p_k^A)\nonumber\\
&&\times\prod_{j=1}^{K_{\bar1}}\frac{x^-(p^B_j)-x_{\bar1,j}}{x^+(p^B_j)-x_{\bar1,j}}\prod_{j=1}^{K_{1}}\frac{1-\frac{1}{x^-(p^B_j)x_{1,j}}}{1-\frac{1}{x^+(p^B_j)x_{1,j}}}
\prod_{j=1}^{K_{\bar3}}\frac{x^-(p^B_j)-x_{\bar3,j}}{x^+(p^B_j)-x_{\bar3,j}}\prod_{j=1}^{K_{3}}\frac{1-\frac{1}{x^-(p^B_j)x_{3,j}}}{1-\frac{1}{x^+(p^B_j)x_{3,j}}}\,,
\\
&&1=\prod_{l=1}^{N_A}\left(\frac{x_{1,j}-x^-(p^A_l)}{x^+_{1,j}-x^+(p^A_l)}\right)
\prod_{l=1}^{N_B}\left(\frac{1-\frac{1}{x_{1,j}x^-(p^B_l)}}{1-\frac{1}{x_{1,j}x^+(p^B_l)}}\right)\,,
\\
&&1=\prod_{l=1}^{N_A}\left(\frac{x_{3,j}-x^-(p^A_l)}{x^+_{3,j}-x^+(p^A_l)}\right)
\prod_{l=1}^{N_B}\left(\frac{1-\frac{1}{x_{3,j}x^-(p^B_l)}}{1-\frac{1}{x_{3,j}x^+(p^B_l)}}\right)\,,
\\
&&1=\prod_{l=1}^{N_B}\left(\frac{x_{\bar1,j}-x^-(p^B_l)}{x^+_{\bar1,j}-x^+(p^B_l)}\right)
\prod_{l=1}^{N_A}\left(\frac{1-\frac{1}{x_{\bar1,j}x^-(p^A_l)}}{1-\frac{1}{x_{\bar1,j}x^+(p^A_l)}}\right)\,,
\\
&&1=\prod_{l=1}^{N_B}\left(\frac{x_{\bar3,j}-x^-(p^B_l)}{x^+_{\bar3,j}-x^+(p^B_l)}\right)
\prod_{l=1}^{N_A}\left(\frac{1-\frac{1}{x_{\bar3,j}x^-(p^A_l)}}{1-\frac{1}{x_{\bar3,j}x^+(p^A_l)}}\right),
\ee
\begin{figure}
\begin{centering}
\includegraphics[height=4cm]{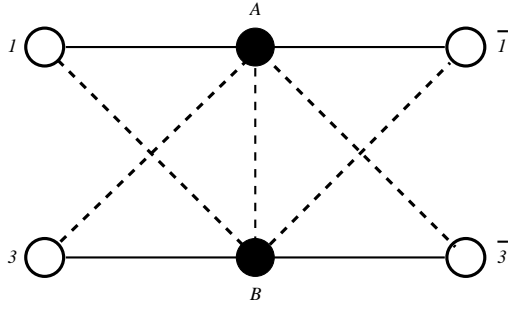}
\par\end{centering}
\caption{${\rm AdS}_3\times S^3\times T^4$:  two momentum-carrying nodes (black dots) are connected to 
four magnonic nodes (circle) after redefinition.}
\label{fig3}
\end{figure}
which match exactly the equations conjectured by  \cite{Babichenko:2009dk, OhlssonSax:2011ms}, if we define $L_{[1,2]}=L+K_A-K_B+K_{\bar1}+K_{\bar3})$.

In analogy with the case of ${\rm AdS}_3\times T^4$, in order to get the Bethe equations written in the notation of \cite{Babichenko:2009dk, OhlssonSax:2011ms}, we need to change $p^{\bar 1, \bar 3}\rightarrow \bar p^{\bar 1, \bar 3}$ in Eqs.(\ref{S3BAEi}-\ref{S3BAEv}) and to redefine
the magnonic nodes corresponding to the variable $2$ into two sets of $2$ and $\bar 2$ variables,
\be
x^{+}(\lambda_{j})=x_{2,j};\quad j=1,\dots,K_{2};\quad x^{+}(\lambda_{j+K_{2}})=1/x_{\bar2,j};\quad j=1,\dots,K_{\bar2},
\ee
as illustrated in Fig.4:
\be
&&\hspace{-0.5cm}e^{ip^{1}_j(L+N_1-N_{\bar1}+K_{\bar2})}=e^{i(P_1-P_{\bar1})}\prod_{k=1,\neq j}^{N_1}\frac{x^+_{1,j}-x^-_{1,k}}{x^-_{1,j}-x^+_{1,k}}\frac{1-\frac{1}{x^+_{1,j} x^-_{1,k}}}{1-\frac{1}{x^-_{1,j} x^+_{1,k}}}
\sigma^2(p_{1,j},p_{1,k})
\prod_{k=1}^{N_{\bar 1}}\sigma^{-2}(p_{1,j},p_{\bar1,k})\nonumber\\
&&\hspace{-0.5cm}\times\prod_{j=1}^{K_2}\left(\frac{x^-(p^1_j)-x_{2,j}}{x^+(p^1_j)-x_{2,j}}\right)\prod_{j=1}^{K_{\bar2}}\left(\frac{1-\frac{1}{x^-(p^1_j)x_{2,j}}}{1-\frac{1}{x^-(p^1_j)x_{2,j}}}\right)\,,\\
&&\hspace{-0.5cm}e^{ip^{3}_j(L+N_3-N_{\bar3}+K_{\bar2})}=e^{i(P_3-P_{\bar3})}\prod_{k=1,\neq j}^{N_3}\frac{x^+_{3,j}-x^-_{3,k}}{x^-_{3,j}-x^+_{3,k}}\frac{1-\frac{1}{x^+_{3,j} x^-_{3,k}}}{1-\frac{1}{x^-_{3,j} x^+_{3,k}}}
\sigma^2(p_{3,j},p_{3,k})
\prod_{k=1}^{N_{\bar 3}}\sigma^{-2}(p_{3,j},p_{\bar3,k})\nonumber\\
&&\hspace{-0.5cm}\times\prod_{j=1}^{K_2}\left(\frac{x^-(p^3_j)-x_{2,j}}{x^+(p^3_j)-x_{2,j}}\right)\prod_{j=1}^{K_{\bar2}}\left(\frac{1-\frac{1}{x^-(p^3_j)x_{2,j}}}{1-\frac{1}{x^-(p^3_j)x_{2,j}}}\right)\,,\\
&&\hspace{-0.5cm}e^{-ip^{\bar1}_j(L+N_{\bar1}-N_1+K_{2})}=e^{i(P_{\bar1}-P_1)}\prod_{k=1,\neq j}^{N_{\bar1}}\frac{x^+_{\bar1,j}-x^-_{\bar1,k}}{x^-_{\bar1,j}-x^+_{\bar1,k}}\frac{1-\frac{1}{x^+_{\bar1,j} x^-_{\bar1,k}}}{1-\frac{1}{x^-_{\bar1,j} x^+_{\bar1,k}}}
\sigma^2(p_{\bar1,j},p_{\bar1,k})
\prod_{k=1}^{N_{1}}\sigma^{-2}(p_{\bar1,j},p_{1,k})\nonumber\\
&&\hspace{-0.5cm}\times\prod_{j=1}^{K_{\bar2}}\left(\frac{x^-(p^{\bar1}_j)-x_{\bar2,j}}{x^+(p^{\bar1}_j)-x_{\bar2,j}}\right)\prod_{j=1}^{K_2}\left(\frac{1-\frac{1}{x^-(p^{\bar1}_j)x_{2,j}}}{1-\frac{1}{x^-(p^{\bar1}_j)x_{2,j}}}\right)\,,
\ee

\be
&&\hspace{-0.5cm}e^{-ip^{\bar3}_j(L+N_{\bar3}-N_3+K_{2})}=e^{i(P_{\bar3}-P_3)}\prod_{k=1,\neq j}^{N_{\bar3}}\frac{x^+_{\bar3,j}-x^-_{\bar3,k}}{x^-_{\bar3,j}-x^+_{\bar3,k}}\frac{1-\frac{1}{x^+_{\bar3,j} x^-_{\bar3,k}}}{1-\frac{1}{x^-_{\bar3,j} x^+_{\bar3,k}}}
\sigma^2(p_{\bar3,j},p_{\bar3,k})
\prod_{k=1}^{N_{3}}\sigma^{-2}(p_{\bar3,j},p_{3,k})\nonumber\\
&&\hspace{-0.5cm}\times\prod_{j=1}^{K_{\bar2}}\left(\frac{x^-(p^{\bar3}_j)-x_{\bar2,j}}{x^+(p^{\bar3}_j)-x_{\bar2,j}}\right)\prod_{j=1}^{K_2}\left(\frac{1-\frac{1}{x^-(p^{\bar3}_j)x_{2,j}}}{1-\frac{1}{x^-(p^{\bar3}_j)x_{2,j}}}\right)\,,\\
&&\hspace{-0.5cm}1=\prod_{\ell=1}^{K_1}\frac{x_{2,j}-x^-(p_{1,\ell})}{x_{2,j}-x^+(p_{1,\ell})}\prod_{\ell=1}^{K_3}\frac{x_{2,j}-x^-(p_{3,\ell})}{x_{2,j}-x^+(p_{3,\ell})}\prod_{\ell=1}^{K_{\bar1}}\frac{1-\frac{1}{x_{2,j}x^-(p_{\bar1,\ell})}}{1-\frac{1}{x_{2,j}x^+(p_{\bar1,\ell})}}\prod_{\ell=1}^{K_{\bar3}}\frac{1-\frac{1}{x_{2,j}x^-(p_{\bar3,\ell})}}{1-\frac{1}{x_{2,j}x^+(p_{\bar3,\ell})}}\,,\\
&&\hspace{-0.5cm}1=\prod_{\ell=1}^{K_{\bar1}}\frac{x_{\bar2,j}-x^-(p_{\bar1,\ell})}{x_{\bar2,j}-x^+(p_{\bar1,\ell})}\prod_{\ell=1}^{K_{\bar3}}\frac{x_{\bar2,j}-x^-(p_{\bar3,\ell})}{x_{\bar2,j}-x^+(p_{\bar3,\ell})}\prod_{\ell=1}^{K_{1}}\frac{1-\frac{1}{x_{\bar2,j}x^-(p_{1,\ell})}}{1-\frac{1}{x_{\bar2,j}x^+(p_{1,\ell})}}\prod_{\ell=1}^{K_{3}}\frac{1-\frac{1}{x_{\bar2,j}x^-(p_{3,\ell})}}{1-\frac{1}{x_{\bar2,j}x^+(p_{3,\ell})}}\,.
\ee
\begin{figure}
\begin{centering}
\includegraphics[height=4cm]{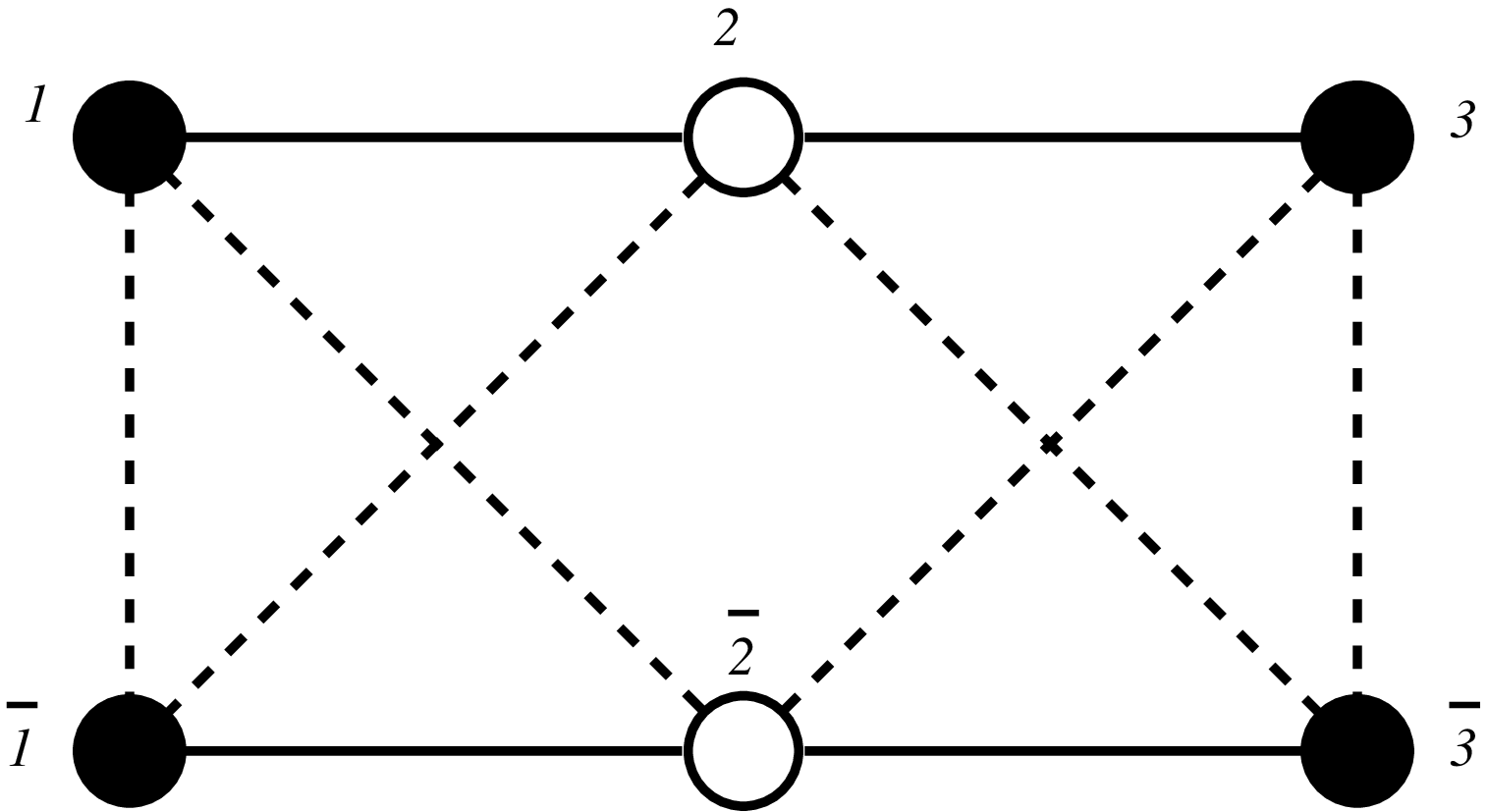}
\par\end{centering}
\caption{${\rm AdS}_3\times S^3\times S^3\times S^1$:  four momentum-carrying nodes (black dots) are 
connected to  two magnonic nodes (circle) after redefinition.}
\label{fig4}
\end{figure}
In this case, to have full agreement with \cite{Babichenko:2009dk, OhlssonSax:2011ms}, we have to redefine the parameter $L$ in different ways in each equation for the momentum-carrying variables:
\be
L_{1}\equiv L+N_1-N_{\bar1}+K_{\bar2}\,;\quad L_{3}\equiv L+N_3-N_{\bar3}+K_{\bar2}\,;\nonumber\\
 L_{\bar1}\equiv L+N_{\bar1}-N_{1}+K_{\bar2}\,;\quad L_{\bar3}\equiv L+N_{\bar3}-N_{3}+K_{\bar2}\,.\nonumber
\ee
This could be useful to solve the apparent disagreement between string results and predictions from the Bethe equations for energies of solutions belonging to the $su(2)\times su(2)$ sector, pointed out by \cite{Rughoonauth:2012qd}, where they would need independent definitions of the spin chain lengths in each $su(2)$ subsector.

\section{Discussion}\label{sec:discussion}

We proposed $su(1|1)\times su(1|1)$- and $su(1|1)$-invariant $S$-matrices for the massive modes of IIB string theory on  ${\rm AdS}_3 \times S^3 \times T^4$ and ${\rm AdS}_3 \times S^3 \times S^3 \times S^1$, respectively. From these we derived the Bethe equations proposed in \cite{Babichenko:2009dk, OhlssonSax:2011ms}. The derivation involved, among other steps, the particle-antiparticle transformation on some momenta of the ``massive'' variables and the doubling of the fermionic variables in a fashion similar to the ${\rm AdS}_5{\rm CFT}_4$ \cite{Martins:2007hb} and ${\rm AdS}_4/{\rm CFT}_3$ \cite{AN} cases.

Some scalar factors remain undetermined and we were able to guess them by requiring the unitarity of the $S$-matrix and the matching with the conjectured BAEs. Because of the appearently missing crossing relations \cite{Janik:2006dc} for the $\mathfrak{su}(1|1)$ algebra \cite{RafEri, David:2010yg}, a more solid derivation of such scalar factors remains as an open problem.

Another open problem is to incorporate the massless modes into the $S$-matrix formulation.
In a relativistic theory, the massless limit can be obtained by shifting the rapidity to $\pm\infty$, which
often makes the $S$-matrices between massive and massless modes trivial.
While this mechanism seems not applicable in our non-relativistic case, we believe a similar argument
may provide a clue.

Albeit these unsolved problems, we believe that our findings can play a r\^ole similar the $su(2|2)$ $S$-matrix \cite{Beisert:2005tm} in both ${\rm AdS}_5/{\rm CFT}_4$ and ${\rm AdS}_4/{\rm CFT}_3$ \cite{AN}, and to lead some deeper understanding of the yet quite unexplored ${\rm AdS}_3/{\rm CFT}_2$.

Finally, it would be interesting to investigate possible exact $S$-matrices for the analogous case of ${\rm AdS}_2/{\rm CFT}_1$, for which a set of all-loop Bethe equations has been recently proposed in \cite{Sorokin:2011rr} \footnote{See also \cite{Zarembo:2010yz} for the derivation of the classical equations in this case and a review about finite-gap integration in various ${\rm AdS}_{d}$ backgrounds.}. 
It will be also interesting to check our proposals in certain perturbative computations. 
One immediate way is to compute
the worldsheet $S$-matrix based on a gauge-fixed string action for the strong coupling limit.
On the other hand, it would be also important to check the reflectionless of our $S$-matrices, as predicted by \cite{David:2010yg}, through some weak coupling perturbative calculations, for example, or along the lines of \cite{AN1}.

\section*{Acknowledgments}
We would like to thank K. Zarembo for useful comments. CA thanks the Centro de F\'isica do Porto (CFP) at the University of Porto for the warm hospitality. DB is grateful to D. Fioravanti, S. Piscaglia and M. Rossi for useful discussions and collaboration on related topics.
This work was supported in part by the WCU Grant No. R32-2008-000-101300 (CA) and the FCT fellowship SFRH/BPD/69813/2010 (DB). CFP is partially funded by FCT through the projects PTDC/FIS/099293/2008 and CERN/FP/116358/2010.

\end{document}